\documentstyle[12pt]{article}
\begin{document}
\pagestyle{empty}
\hspace*{12.4cm}IU-MSTP/39 \\
\hspace*{13cm}hep-lat/0001029 \\
\hspace*{13cm}January, 2000
\begin{center}
 {\Large\bf Topological Charge of Lattice Abelian Gauge Theory}
\end{center}

\vspace*{1cm}
\def\thefootnote{\fnsymbol{footnote}}
\begin{center}{\sc Takanori Fujiwara${}^1$,} 
{\sc Hiroshi Suzuki${}^2$}\footnote{Permanent address: 
Department of Mathematical Sciences, Ibaraki University,
Mito 310-8512, Japan}
and {\sc Ke Wu${}^1$}\footnote{Permanent address: 
Institute of Theoretical Physics, Academia Sinica, P.O.Box 
2735, Beijing 
100080, China}
\end{center}
\vspace*{0.2cm}
\begin{center}
{\it ${}^1$Department of Mathematical Sciences, Ibaraki 
University,
Mito 310-8512, Japan \par
${}^2$ High Energy Group, Abdus Salam ICTP, Trieste, 34014, 
Italy}
\end{center}
\vfill
\begin{center}
{\large\sc Abstract}
\end{center}
\noindent
Configuration space of abelian gauge theory on a periodic lattice 
becomes topologically disconnected by excising exceptional gauge 
field configurations. It is possible to define a $U(1)$ bundle 
from the nonexceptional link variables by a smooth interpolation 
of the transition functions. The lattice analogue of Chern character 
obtained by a cohomological technique based on the noncommutative 
differential calculus is shown to give a topological charge related 
to the topological winding number of the $U(1)$ bundle. 


\newpage
\pagestyle{plain}

\section{Introduction}
\label{sec:intro}
\setcounter{equation}{0}

In a recent paper \cite{lus3} L\"uscher has investigated generic 
structures of chiral anomaly for abelian gauge theory on the 
lattice. 
His work is extended to arbitrary higher dimensions in our previous 
papers \cite{fswPLB463}, where the topological part of the 
axial anomaly is shown to be interpretable as a lattice 
generalization of 
Chern character within the framework of noncommutative differential 
calculus \cite{sit}. In the continuum theory the Chern character 
gives the 
integer topological winding number when integrated over the base 
manifold and it coincides with the index of the Dirac 
operator \cite{AS}. 
The chiral anomaly on the lattice can also be related with 
the index \cite{HLN} of the Ginsparg-Wilson Dirac 
operator \cite{GW,has}. 
So it is very natural to expect some extension from the continuum 
to the lattice of the index theorem relating the analytical index 
of the Dirac 
operator with the topological invariant of the manifold on which 
the Dirac operator is defined . In this respect, it is, however, 
not clear in the constructions of ref. \cite{lus3,fswPLB463} 
whether the lattice analogue of Chern character can be related to 
some topology of the gauge theory on the lattice. 

One might think that it would make no sense to argue 
the topological configurations on the lattice since any 
lattice fields could be 
continuously deformed into the trivial configuration and no 
nontrivial topological invariants could be constructed. 
But this is not the case. As argued in 
refs. \cite{lus4,phil,PS,GKLSW}, 
it is indeed possible to define a smooth fiber bundle and hence a 
topological winding number for a given lattice field configuration 
if it contains no exceptional link variables \cite{lus4,phil,GKLSW}. 
In the case of abelian theories L\"uscher has shown 
in ref. \cite{lus5} that the configuration space of the link 
variables 
satisfying the admissibility condition has a rich topological 
structure. 
It is considered as a kind of smoothness condition for the gauge 
field 
configuration, ensuring the existence of gauge potentials 
continuously 
parameterizing the link variables. The essential point here is that 
the 
configuration space of the admissible link variables is 
topologically 
disconnected.

In this paper we investigate the topological charge of abelian gauge 
theory on a periodic lattice in arbitrary even dimensions and argue 
that the lattice analogue of Chern character obtained in refs. 
\cite{lus3,fswPLB463} indeed gives an integer-valued topological 
invariant by relating it to the topological winding 
number of a $U(1)$ bundle constructed from the lattice gauge fields 
by the interpolation method of refs. \cite{lus4,GKLSW}.

We should add a brief argument concerning the 
theorem given in refs. \cite{lus3,fswPLB463}, 
where it is supposed for an infinite hypercubic regular lattice. 
We can extend it to topologically nontrivial lattices $\Lambda$ 
without boundaries by restricting to ultralocal functions. We assume 
that $\Lambda$ is locally hypercubic and regular. This means that
for any point $n\in\Lambda$ one can find a set $U_n$ of lattice 
points 
and links with a hypercubic regular lattice structure of the same 
dimensions.
Hypercubic regular lattices with periodic boundary conditions, 
which we shall consider, are examples for $\Lambda$. We call 
functions $f$ on $\Lambda$ ultralocal if $f(n)$ for any 
$n\in \Lambda$ depends only on the gauge potentials associated 
to links within the subset $U_n$ of $\Lambda$. The abelian gauge 
potentials on the lattice will be treated in Sect. 3 in detail. 
See also refs. \cite{thooft,lus3,lus5}.
Throughout this paper we assume the lattice spacing $a=1$. 
The forward and backward difference operators ${\it \Delta}_\mu$ and 
${\it \Delta}^\ast_\mu$ are then defined by 
\begin{eqnarray}
  \label{eq:fbdiff}
  {\it \Delta}_\mu f(n)=f(n+\hat\mu)-f(n)~, \qquad
  {\it \Delta}^\ast_\mu f(n)=f(n)-f(n-\hat\mu)~.
\end{eqnarray}
Then the theorem is extended to the lattice $\Lambda$ as 

\vskip .3cm
\noindent{\bf Theorem :} {\sl Let $q$ be gauge invariant and smooth 
ultralocal function of the abelian gauge potentials $A_\mu$ on a 
locally 
hypercubic regular lattice $\Lambda$ of dimensions $D$ without 
boundaries that 
satisfies the topological invariance 
\begin{eqnarray}
  \label{eq:topinv}
  \sum_{n\in\Lambda}\delta q(n)&=&0
\end{eqnarray}
for arbitrary local variations of the gauge potentials 
$A_\mu\rightarrow A_\mu+\delta A_\mu$, then $q(n)$ for arbitrary 
$n\in\Lambda$ takes the form 
\begin{eqnarray}
  \label{eq:theorem}
  q(n)&=&\sum_{l=0}^{[D/2]}\beta_{\mu_1\nu_1\cdots\mu_l\nu_l}
  F_{\mu_1\nu_1}(n)
  F_{\mu_2\nu_2}(n+\mu_1+\nu_1) \nonumber\\
  &&\times\cdots\times 
  F_{\mu_l\nu_l}(n+\mu_1+\nu_1+\cdots+\mu_{l-1}+\nu_{l-1})
  +{\it \Delta}^\ast_\mu k_\mu(n)~,
\end{eqnarray}
where $F_{\mu\nu}(n)={\it \Delta}_\mu A_\nu(n)-{\it \Delta}_\nu 
A_\mu(n)$ is the 
field strength, the coefficient 
$\beta_{\mu_1\nu_1\cdots\mu_n\nu_n}$ is 
antisymmetric in its indices and the current $k_\mu$ can be chosen 
to be 
gauge invariant and ultralocal in the gauge potential.}
\vskip .3cm
For functions $q$ on the infinite lattice ${\bf Z}^D$ the theorem 
holds 
true. Since $\Lambda$ is assumed to be locally hypercubic and 
regular, 
the same identity should also follow for ultralocal functions $q$. 
The 
point here is that the gauge invariant current $k_\mu$ can be 
chosen to 
be ultralocal since both $q$ and the topological terms are 
ultralocal 
and (\ref{eq:theorem}) must be an identity for any configuration of 
the 
gauge potentials. As we will see, the topological charge density in 
general has a complicated form. The theorem will be used to rewrite 
it to the standard form (\ref{eq:theorem}). 

This paper is organized as follows. In the next section we describe 
the topological winding number of abelian gauge theory on a $D$ 
dimensional 
torus, giving an explicit formula for the topological charge in 
terms 
of the transition functions. Sect. 3 deals with abelian gauge fields
on a $D$ dimensional periodic lattice. We argue the precise relation 
among the link variables, the field strengths and the gauge 
potentials 
on the lattice. In Sect. 4 we give a formulation of interpolation 
of the parallel transport functions from the discrete lattice to the 
continuum. The interpolated transition functions are obtained in 
closed form. In Sect. 5 we define the topological charge of the 
lattice abelian gauge theory by the Chern number of the fiber bundle 
with the interpolated transition functions given in Sect. 4. The 
connection between the topological charge and the chiral 
anomaly obtained by the method of noncommutative differential 
calculus in our previous works is established. We also show 
that the topological charge can be expressed solely by the 
magnetic fluxes introduced in ref. \cite{lus5}. A mathematical 
lemma concerning the existence of periodic potential functions 
for flux-free field configurations is given in Sect. 6. In the 
proof of the lemma we show the systematic way to isolate 
the flux contributions from the field strengths. For illustration 
we present a simple but nontrivial field configuration of constant 
field strengths with unit topological charges in two 
dimensions. Sect. 7 is devoted to summary and discussion.

\section{Topological charge of abelian gauge theory on $T^D$}
\label{sec:tcagttd}
\setcounter{equation}{0}

Fiber bundles over a manifold are topologically classified 
by the equivalence class of transition functions. Our first main 
concern 
is to give a formula for the topological winding number of the fiber
bundle in terms of transition functions. In this paper we consider 
$U(1)$ 
bundles over a torus $T^D$ of dimensions $D=2N$ defined by the 
identification 
\begin{eqnarray}
  \label{eq:torus}
  x\sim x+L\hat\mu \quad \hbox{for}\quad
  x\in{\bf R}^D~,~\mu=1,\cdots,D
\end{eqnarray}
where $\hat\mu$ denotes the unit vector in the $\mu$-th 
direction and the period $L$ of the torus is assumed to be 
a positive integer. A hypercubic periodic lattice $\Lambda$ 
of dimensions $D$ is defined as the set of integral lattice 
points in $T^D$. 

We divide $T^D$ into $L^D$ hypercubes $c(n)$ 
($n\in\Lambda$) defined by
\begin{eqnarray}
  \label{eq:c(n)}
  c(n)=\{x\in T^D|x=n
  +\sum_{\mu=1}^Dy_\mu\hat\mu~,~0\le y_\mu\le1\}~.
\end{eqnarray}
We assume that $L$ is large enough so that any 
restricted bundle over 
$c(n)$ is trivial. Mathematically this can be achieved for $L\ge2$.
For later convenience, let us denote the intersection of $c(n)$ and 
$c(n-\hat\mu)$ by $p(n,\mu)$ and the common boundary of $p(n,\mu)$, 
$p(n,\nu)$, $\cdots$, $p(n,\sigma)$ by $p(n,\mu,\nu,\cdots,\sigma)$.

Let $A^{(n)}(x)$ be the gauge potential 1-form\footnote{The gauge 
field 
$A^{(n)}_\mu(x)$ is assumed to be real.} on the hypercube $c(n)$,
then the gauge potentials $A^{(n-\hat\mu)}$ and $A^{(n)}$ are 
related 
by a gauge transformation on $p(n,\mu)$
\begin{eqnarray}
  \label{eq:1}
  A^{(n-\hat\mu)}(x)&=&A^{(n)}(x)+d\Lambda_{n,\mu}(x)~,
\end{eqnarray}
where $d$ is the ordinary exterior differential and 
$\Lambda_{n,\mu}$ is defined by the transition functions 
$v_{n,\mu}$ as 
\begin{eqnarray}
  \label{eq:15}
  v_{n,\mu}={\rm e}^{-i\Lambda_{n,\mu}}~.
\end{eqnarray}
The transition functions must satisfy cocycle conditions 
for $x\in p(n,\mu,\nu)$ \cite{lus4}
\begin{eqnarray}
 \label{eq:cocycle}
  v_{n-\hat\mu,\nu}(x)v_{n,\mu}(x)=v_{n-\hat\nu,\mu}(x)v_{n,\nu}(x)~.
\end{eqnarray}
In terms of $\Lambda_{n,\mu}$ they can be written as
\begin{eqnarray}
  \label{eq:4}
  {\it \Delta}^\ast_\mu\Lambda_{n,\nu}
  ={\it \Delta}^\ast_\nu\Lambda_{n,\mu}
  \qquad (\hbox{mod}~ 2\pi)~,
\end{eqnarray}
where ${\it \Delta}^\ast_\mu$ is the backward difference operator 
defined by 
(\ref{eq:fbdiff}).

The topological charge of the abelian gauge theory is given by
\begin{eqnarray}
  \label{eq:5}
  Q=C_N\int_{T^D} F^N~,\qquad C_N=\frac{1}{(2\pi)^NN!}~,
\end{eqnarray}
where $F(x)=dA^{(n)}(x)$ is the field strength 2-form for $x\in 
c(n)$. It is gauge invariant and globally defined on $T^D$.  
By the Bianchi identity $dF=0$, we have $F^N=d(A^{(n)}F^{N-1})$ 
and hence 
\begin{eqnarray}
  \label{eq:6}
  Q&=&C_N\sum_{n\in\Lambda}\int_{c(n)} d(A^{(n)}F^{N-1}) \nonumber \\
  &=&C_N\sum_{n\in\Lambda}\sum_\mu\Biggl(\int_{p(n+\mu,\mu)}
  A^{(n)}F^{N-1}
  -\int_{p(n,\mu)}A^{(n)}F^{N-1}\Biggr) \nonumber \\
  &=&C_N\sum_{n\in\Lambda}\sum_\mu\int_{p(n,\mu)}
  (A^{(n-\hat\mu)}-A^{(n)})F^{N-1} \nonumber \\
  &=&C_N\sum_{n\in\Lambda}\sum_\mu\int_{p(n,\mu)}
  d\Lambda_{n,\mu}F^{N-1}~.
\end{eqnarray}
In the last step we have used the relation (\ref{eq:1}). Since 
$d\Lambda_{n,\mu}$ is a closed form, we  can write 
$d\Lambda_{n,\mu}F^{N-1}
=-d(d\Lambda_{n,\mu}A^{(n)}F^{N-2})$ and $Q$ is further reduced as
\begin{eqnarray}
  \label{eq:7}
  Q&=&-C_N\sum_n\sum^\sim_{\mu,\nu}\int_{p(n,\mu,\nu)}
  \biggl\{-{\it \Delta}^\ast_\nu d\Lambda_{n,\mu}
  A^{(n)}F^{N-2}+d\Lambda_{n-\hat\nu,\mu}d\Lambda_{n,\nu}
  F^{N-2}\biggr\}~,
\end{eqnarray}
where $\displaystyle{\sum^\sim_{\mu,\nu, \cdots}}$ implies 
antisymmetrized summation on $\mu,\nu,\cdots$ satisfying
\begin{eqnarray}
  \label{eq:10}
  \sum^\sim_{\mu,\nu,\cdots}f_{\mu\nu\cdots}
  =f_{12\cdots}+\cdots~, \qquad
  \sum^\sim_{\cdots,\mu_i, \cdots,\mu_j\cdots}=
  -\sum^\sim_{\cdots,\mu_j, \cdots,\mu_i\cdots}~.
\end{eqnarray}

Since $d$ commutes with ${\it \Delta}^\ast_\mu$ and 
$d{\it \Delta}^\ast_\nu
\Lambda_{n,\mu}$ is symmetric in $\mu$ and $\nu$ by the cocycle 
condition (\ref{eq:4}), the first term in the integrand of 
(\ref{eq:7}) 
does not contribute to $Q$. We thus obtain
\begin{eqnarray}
  \label{eq:8}
  Q&=&-C_N\sum_n\sum^\sim_{\mu,\nu}\int_{p(n,\mu,\nu)}
  d\Lambda_{n-\hat\nu,\mu}d\Lambda_{n,\nu}
  F^{N-2}~.
\end{eqnarray}
Similar procedure can be repeated until all the field strengths 
disappear. The final expression for $Q$ is given by
\begin{eqnarray}
  \label{eq:9}
  Q&=&(-1)^{\frac{N(N-1)}{2}}C_N\sum_n\sum^\sim_{\mu_1,\cdots,\mu_N}
  \int_{p(n,\mu_1,\mu_2,\cdots,\mu_N)}
  d\Lambda_{n-\hat\mu_2-\cdots-\hat\mu_N,\mu_1}
  d\Lambda_{n-\hat\mu_3-\cdots-\hat\mu_N,\mu_2} 
  \cdots d\Lambda_{n,\mu_N} \nonumber \\
  &=&(-1)^{\frac{N(N-1)}{2}}C_N\sum_n
  \epsilon_{\mu_1\mu_2\cdots\mu_D}
  \int_{p(n,\mu_1,\mu_2,\cdots,\mu_N)}d^Nx
  \partial_{\mu_{N+1}}\Lambda_{n-\hat\mu_2-\cdots-\hat\mu_N,\mu_1} 
  \partial_{\mu_{N+2}}\Lambda_{n-\hat\mu_3-\cdots-\hat\mu_N,\mu_2} 
  \nonumber \\
  &&\hskip 9cm \times\cdots\times \partial_{\mu_D}\Lambda_{n,\mu_N}~,
\end{eqnarray}
where $\epsilon_{\mu_1\cdots\mu_D}$ is the Levi-Civita symbol 
in $D$ dimensions, $\partial_\mu$ stands for the derivative with 
respect to $x_\mu$, and $d^Nx$ is the volume form on 
$p(n,\mu_1,\cdots,\mu_N)$ . 

To show that no new condition other than the cocycle conditions 
(\ref{eq:4}) is necessary we prove it by mathematical induction. 
Let us assume that 
\begin{eqnarray}
  \label{eq:11}
  Q&=&(-1)^{\frac{k(k-1)}{2}}C_N\sum_n\sum^\sim_{\mu_1,\cdots,\mu_k}
  \int_{p(n,\mu_1,\mu_2,\cdots,\mu_k)}
  d\Lambda_{n-\hat\mu_2-\cdots-\hat\mu_k,\mu_1}
  d\Lambda_{n-\hat\mu_3-\cdots-\hat\mu_k,\mu_2} \nonumber \\
  &&\hskip 8cm \times 
  \cdots \times d\Lambda_{n,\mu_k}F^{N-k}
\end{eqnarray}
holds true up to some integer $k<N$. Then by carrying out the 
manipulation 
from (\ref{eq:6}) to (\ref{eq:7}) we have
\begin{eqnarray}
  \label{eq:12}
  Q&=&(-1)^{\frac{k(k+1)}{2}}C_N\sum_n\sum^\sim_{\mu_1,\cdots,
    \mu_{k+1}}
  \int_{p(n,\mu_1,\mu_2,\cdots,\mu_k,\mu_{k+1})} \nonumber \\
  &&\times\biggl(-{\it \Delta}^\ast_{\mu_{k+1}}
  (d\Lambda_{n-\hat\mu_2-\cdots-\hat\mu_k,\mu_1}
  d\Lambda_{n-\hat\mu_3-\cdots-\hat\mu_k,\mu_2} 
  \cdots d\Lambda_{n,\mu_k})A^{(n)} \nonumber \\
  &&+d\Lambda_{n-\hat\mu_2-\cdots-\hat\mu_k-\hat\mu_{k+1},\mu_1}
  d\Lambda_{n-\hat\mu_3-\cdots-\hat\mu_k-\hat\mu_{k+1},\mu_2} 
  \cdots d\Lambda_{n-\hat\mu_{k+1},\mu_k}d\Lambda_{n,\mu_{k+1}}
  \biggr)F^{N-k-1}~.
\nonumber \\
\end{eqnarray}
By noting the Leibnitz rule on the lattice
${\it \Delta}^\ast_\mu (f_n g_n)={\it \Delta}^\ast_\mu f_ng_n+f_{n-\hat\mu}
{\it \Delta}^\ast_\mu g_n$, we get 
\begin{eqnarray}
  \label{eq:13}
  &&{\it \Delta}^\ast_{\mu_{k+1}}
  (d\Lambda_{n-\hat\mu_2-\cdots-\hat\mu_k,\mu_1}
  d\Lambda_{n-\hat\mu_3-\cdots-\hat\mu_k,\mu_2} 
  \cdots d\Lambda_{n,\mu_k})\nonumber \\
  &&\hskip 1cm =d{\it \Delta}^\ast_{\mu_{k+1}}
  \Lambda_{n-\hat\mu_2-\cdots-\hat\mu_k,\mu_1} 
  d\Lambda_{n-\hat\mu_3-\cdots-\hat\mu_k,\mu_2}
  \cdots d\Lambda_{n,\mu_k} \nonumber \\
  &&\hskip 1.5cm
  +d\Lambda_{n-\hat\mu_2-\cdots-\hat\mu_k-\hat\mu_{k+1},\mu_1}
  d{\it \Delta}^\ast_{\mu_{k+1}}\Lambda_{n-\hat\mu_3-\cdots
    -\hat\mu_k,\mu_2} 
  \cdots d\Lambda_{n,\mu_k}\nonumber \\
  &&\hskip 1.5cm+\cdots+d\Lambda_{n-\hat\mu_2-\cdots
    -\hat\mu_k-\hat\mu_{k+1},\mu_1}
  d\Lambda_{n-\hat\mu_3-\cdots-\hat\mu_k-\hat\mu_{k+1},\mu_2} 
   \cdots d{\it \Delta}^\ast_{\mu_{k+1}}\Lambda_{n,\mu_k}~.
\end{eqnarray}
The $j$-th term of the rhs of this expression is symmetric in 
$\mu_j$ and $\mu_{k+1}$ for $j=1,\cdots,k$ by the cocycle condition 
(\ref{eq:4}), hence the 
first term of the integrand of (\ref{eq:12}) does not contribute 
to the 
topological charge due to the antisymmetrized sum. The resulting 
expression 
for $Q$ is just (\ref{eq:11}) with $k$ replaced by $k+1$, implying 
that 
(\ref{eq:11}) holds true for any $k\le N$. This completes the proof 
of (\ref{eq:9}).

\section{Lattice abelian gauge field on the periodic lattice}
\label{sec:lagf}
\setcounter{equation}{0}

The link variables $U_\mu(n)$ ($n\in\Lambda,~\mu=1,\cdots,D$) 
are subject to the periodic boundary conditions
\begin{eqnarray}
  \label{eq:x1}
  U_\mu(n+L\hat\nu)=U_\mu(n)~,\qquad (\nu=1,\cdots,D)~, 
\end{eqnarray} 
and are assumed to be parametrized by 
\begin{eqnarray}
  \label{eq:upba}
  U_\mu(n)&=&{\rm e}^{i a_\mu(n)} ~, \qquad (-\pi\le a_\mu(n)<\pi)~,
\end{eqnarray}
where $a_\mu(n)$ is a vector field on $\Lambda$. The field strength 
is defined by 
\begin{eqnarray}
  \label{eq:fieldstr}
  F_{\mu\nu}(n)=\frac{1}{i}\ln U_\mu(n)U_\nu(n+\hat\mu)
  U_\mu(n+\hat\nu)^{-1}U_\nu(n)^{-1}~,\qquad
  |F_{\mu\nu}(n)|<\pi~.
\end{eqnarray}
We exclude the exceptional field configurations  \cite{phil} 
with $|F_{\mu\nu}(n)|=\pi$, 
where the plaquette variable equals $-1$ and $(-1)^y$ ($0\le y\le1$) 
becomes ambiguous. 

The field strengths can be written in terms of $a_\mu$ as
\begin{eqnarray}
  \label{eq:ftoa}
  F_{\mu\nu}(n)={\it \Delta}_\mu a_\nu(n)-{\it \Delta}_\nu a_\mu(n)
  +2\pi n_{\mu\nu}(n)~,
\end{eqnarray}
where ${\it \Delta}_\mu$ is the forward difference operator and 
$n_{\mu\nu}$ is an integer-valued anti-symmetric tensor field on 
$\Lambda$. 
The $n_{\mu\nu}(n)$ must be chosen so that the field strength 
$F_{\mu\nu}(n)$ lies within the principal branch of the 
logarithm in (\ref{eq:fieldstr}). Obviously, it satisfies 
$|n_{\mu\nu}(n)|\le2$. Furthermore, the link variables are 
assumed to be restricted so that the field strengths always 
satisfy the Bianchi identity ${\it \Delta}_{[\lambda}
F_{\mu\nu]}(n)=0$ in order to ensure the existence of gauge 
potential $A_\mu$ as 
\begin{eqnarray}
  \label{eq:gaugepot}
  F_{\mu\nu}(n)&=&{\it \Delta}_\mu A_\nu(n)-{\it \Delta}_\nu 
  A_\mu(n)~.
\end{eqnarray}
This requirement is automatically satisfied by the admissibility 
condition of ref. \cite{lus3,lus5} that the field strengths 
satisfy
\begin{eqnarray}
  \label{eq:admissi}
  \sup_{n,\mu,\nu}|F_{\mu\nu}(n)|&<&\epsilon  
\end{eqnarray}
for a fixed constant $0<\epsilon<\pi/3$.\footnote{As noted in ref. 
\cite{lus3}, $\epsilon$ should be replaced with $\epsilon/a^2$ for 
the lattice spacing $a\ne1$ and, hence, the restriction on the 
gauge field configuration disappears in the classical 
continuum limit.}
The exceptional filed configurations mentioned above are 
also excluded by this condition. 

Gauge transformations on $\Lambda$ are defined by 
\begin{eqnarray}
  \label{eq:gtrans}
  U_\mu(n)&\rightarrow& V(n)U_\mu(n)V(n+\hat\mu)^{-1}~,
\end{eqnarray}
where $V$ is a $U(1)$-valued function on $\Lambda$. The field 
strengths (\ref{eq:fieldstr}) is obviously gauge invariant. 
Note, however, that neither ${\it \Delta}_\mu a_\nu(n)
-{\it \Delta}_\nu a_\mu(n)$ nor $n_{\mu\nu}(n)$ are separately 
gauge invariant. To see this we parameterize the gauge 
transformations $V$ by a function $\lambda$ on $\Lambda$ as 
$V(n)={\rm e}^{i\lambda(n)}$, where $\lambda$ is assumed 
to satisfy $\lambda(n+L\hat\mu)=\lambda(n)$ and 
$|\lambda(n)|\le\pi$. Then $a_\mu(n)$ and $n_{\mu\nu}(n)$
are transformed as
\begin{eqnarray}
  \label{eq:angtr}
  a_\mu(n)\rightarrow a_\mu(n)-{\it \Delta}_\mu\lambda(n)
  +2\pi N_\mu(n)~, \quad
  n_{\mu\nu}(n)\rightarrow n_{\mu\nu}(n)-{\it \Delta}_\mu N_\nu(n)
  +{\it \Delta}_\nu N_\mu(n)~,
\end{eqnarray}
where $N_\mu$ is an integer-valued vector field on $\Lambda$ 
and must be chosen to satisfy 
\begin{eqnarray}
  \label{eq:Nmu}
  -\pi\le a_\mu(n)-{\it \Delta}_\mu\lambda(n)+2\pi N_\mu(n)<\pi~.
\end{eqnarray}

If the field strength satisfies the Bianchi identities, so does
$n_{\mu\nu}$. Hence it is always possible to find an 
integer-valued vector field $m_\mu$ on ${\bf Z}^D$ satisfying 
\begin{eqnarray}
  \label{eq:dmmu}
  A_\mu(n)=a_\mu(n)+2\pi m_\mu(n)~, \qquad 
  {\it \Delta}_\mu m_\nu(n)-{\it \Delta}_\nu m_\mu(n)=n_{\mu\nu}(n)~.
\end{eqnarray}
To show this we note that $m_\mu(n)$ is only determined up to 
integer-valued gauge transformations $m_\mu(n)
\rightarrow m_\mu(n)-{\it \Delta}_\mu{\it \Lambda}(n)$ 
(${\it \Lambda}(n)\in{\bf Z}$)
on ${\bf Z}^D$ 
and we can always work in the axial gauge $m_D(n)=0$. In this gauge 
$m_\mu(n)$ ($\mu\ne D$) satisfies 
${\it \Delta}_Dm_\mu(n)=-n_{\mu D}(n)$, 
which can be integrated to  
\begin{eqnarray}
  \label{eq:mmusol1}
  m_\mu(n)&=&-\sum_{y_D=0}^{n_D-1}n_{\mu D}(n_1,\cdots,n_{D-1},y_D)
  +m_\mu(n_1,\cdots,n_{D-1},0)~.
\end{eqnarray}
The sum on $y_D$ in the rhs must make sense for arbitrary 
integer $n_D$. 
This can be achieved by defining the sum as
\begin{eqnarray}
  \label{eq:esum}
  \sum_{y=a}^{b-1}f(y)=\cases{
    f(a)+\cdots+f(b-1) & $(b>a)$ \cr
    0                  & $(b=a)$ \cr
    -f(b)-\cdots-f(a-1) & $(b<a)$}
\end{eqnarray}
for arbitrary functions $f$ on ${\bf Z}$ and $a,b\in{\bf Z}$. 
It is just an analogue of one-dimensional integral on the discrete 
space ${\bf Z}$ and satisfies the following properties 
\begin{eqnarray}
  \label{eq:sumprop}
  && \sum_{y=a}^{b-1}f(y)=-\sum_{y=b}^{a-1}f(y)~, \qquad
  \sum_{y=a}^{b-1}f(y)=\sum_{y=a}^{c-1}f(y)+\sum_{y=c}^{b-1}f(y)~,
  \nonumber\\
  && {\it \Delta}_x\Biggl(\sum_{y=a}^{x-1}f(y)\Biggr)=f(x)~, \qquad
  \sum_{y=a}^{b-1}{\it \Delta}_yf(y)=f(b)-f(a)~,
\end{eqnarray}
where ${\it \Delta}_z$ stands for the difference operator with 
respect to 
$z\in{\bf Z}$. In (\ref{eq:mmusol1}) $m_\mu(n)|_{n_D=0}
=m_\mu(n_1,\cdots,n_{D-1},0)$ ($\mu\ne D$) are still to be 
determined. 
In order for (\ref{eq:dmmu}) to be consistent they must satisfy 
the following set of equations
\begin{eqnarray}
  \label{eq:md-1}
  {\it \Delta}_\mu m_\nu(n_1,\cdots,n_{D-1},0)
  -{\it \Delta}_\nu m_\mu(n_1,\cdots,n_{D-1},0)
  =n_{\mu\nu}(n_1,\cdots,n_{D-1},0)~.
\end{eqnarray}
We see that the problem of finding $m_\mu(n)$ in $D$ dimensions 
is reduced to the problem of solving the original equations 
dimensionally reduced to $D-1$ dimensions. Again we can choose the 
axial gauge $m_{D-1}(n_1,\cdots,n_{D-1},0)=0$ and solve 
(\ref{eq:md-1}) 
as before. Obviously, this reduction process can be continued until 
all the $m_\mu$'s are completely determined. We thus obtain 
\begin{eqnarray}
  \label{eq:mmusol}
  m_\mu(n)&=&-\sum_{\nu>\mu}\sum_{y_\nu=0}^{n_\nu-1}
  n_{\mu\nu}(n_1,\cdots,n_{\nu-1},y_\nu,0,\cdots,0)~.
\end{eqnarray}
It is integer-valued as announced. 

From (\ref{eq:dmmu}) the gauge potential $A_\mu$ 
also serves as the local coordinates for the link variables by the 
relations $U_\mu(n)={\rm e}^{iA_\mu(n)}$ \cite{lus3,lus5} and the 
gauge 
transformation (\ref{eq:angtr}) simply becomes
\begin{eqnarray}
  \label{eq:gtraa}
  A_\mu(n)&\rightarrow& A_\mu(n)-{\it \Delta}_\mu\lambda(n)
  -2\pi{\it \Delta}_\mu{\it \Lambda(n)}~,
\end{eqnarray}
where ${\it \Lambda}$ is an arbitrary integer-valued function on 
${\bf Z}^D$. We emphasize that the gauge potential $A_\mu(n)$ 
defined by (\ref{eq:dmmu}) and (\ref{eq:mmusol}) is continuous 
at $U_\mu(n)=-1$ though $a_\mu(n)$ exhibits a discontinuity. 

Note that the periodicity of the link variables and the field 
strengths only implies that of the gauge potentials up to 
gauge trasnformations by $2\pi{\it \Lambda}$ with 
${\it \Lambda}(n)\in {\bf Z}$.
In order to obtain nonvanishing topological charge it is necessary 
to have nonperiodic gauge potentials. For $m_\mu(n)$ given by 
(\ref{eq:mmusol}) the periodicity of $A_\mu(n)$ is completely 
determined by $n_{\mu\nu}(n)$ as 
\begin{eqnarray}
  \label{eq:aperio}
  A_\mu(n+L\hat\nu)-A_\mu(n)&=&2\pi m_\mu(n+L\hat\nu)
  -2\pi m_\mu(n) \nonumber\\
  &=&\cases{\displaystyle{-2\pi\sum_{n_\nu=0}^{L-1}}
    n_{\mu\nu}(n_1,\cdots,n_\nu,0,\cdots,0)
      & for $\mu<\nu$ \cr
    0 & for $\mu\ge\nu$ }
\end{eqnarray}

\section{Interpolated transition functions}
\label{sec:itf}
\setcounter{equation}{0}

We now turn to the construction of the transition functions from the 
gauge fields on the lattice $\Lambda$ by using the interpolation 
technique given in refs. \cite{lus4,GKLSW}. 
The main concerns of the authors 
of refs. \cite{lus4,PS,GKLSW} were 
the construction of the topological charges 
for nonabelian theories in four dimensions. 
The interpolated transition 
functions become more and more complicated 
as the dimensions increase in 
the nonabelian case so that the general 
expression cannot be available
in arbitrary dimensions. In the abelian case, however, all the 
complications related to the noncommutativity of the transition 
functions\footnote{In the nonabelian theories 
the transition functions 
are in fact matrix-valued in some representation.} disappear and it 
is possible to give transition functions in 
closed form as we will show.

We first define parallel transport functions $w^n(\bar x)$ by
\begin{eqnarray}
  \label{eq:x3}
  w^n(\bar x)=U_1(n)^{\sigma_1}U_2(n+\sigma_1\hat 1)^{\sigma_2}
  \cdots U_D(n+\sigma_1\hat1+\sigma_2\hat2+\cdots
  +\sigma_{D-1}\widehat{D-1})^{\sigma_D}
\end{eqnarray}
for points $\bar x$ of the corners of $c(n)$ given by
\begin{eqnarray}
  \label{eq:x2}
  \bar x=n+\sum_{\mu=1}^D \sigma_\mu\hat\mu~, \qquad 
  (\sigma_\mu=\{0,1\})~.
\end{eqnarray}
At this point they are only defined on the corners of $c(n)$. 
In what 
follows we define iteratively interpolations 
of $w^n(\bar x)$ over the 
boundary $\partial c(n)$. 

Before going into mathematical detail, we mention the subtleties in 
defining arbitrary exponents $(w^n(\bar x))^y$ ($0\le y\le1$). 
We define exponents for link variables by $(U_\mu(n))^y
\equiv{\rm e}^{iya_\mu(n)}$. 
But $(w^n(\bar x))^y$ does not coincide with 
$(U_1(n)^{\sigma_1})^y\cdots (U_D(n+\sigma_1\hat1+\cdots
+\sigma_{D-1}\widehat{D-1})^{\sigma_D})^y$ in general.  
The subtlety concerning the 
violation of such a naive distribution law of 
exponents can be avoided if we assume that 
$|a_\mu(n)|$ is sufficiently 
small for any link variable so that the naive distribution law is 
applicable. We assume this for the time being to justify 
the mathematical manipulations and remove such restrictions from the 
final expression of the transition functions. 

Let $\mu_1,\cdots,\mu_{D-1}=\{1,\cdots,D\}\backslash \{\mu\}$ be the 
$D-1$ indices satisfying $\mu_1<\cdots<\mu_{D-1}$. 
Then the interpolation 
along the $\hat\mu_{D-1}$ direction is defined by 
\begin{eqnarray}
  \label{eq:x4}
  w^m\Biggl(n+\sum_{k=1}^{D-2}\sigma_k\hat\mu_k
  +y_{D-1}\hat\mu_{D-1}\Biggr)
  &\equiv& w^m\Biggl(n+\sum_{k=1}^{D-2}
  \sigma_k\hat\mu_k+\hat\mu_{D-1}
  \Biggr)^{y_{D-1}} \nonumber \\
  &&\times w^m\Biggl(n+\sum_{k=1}^{D-2}
  \sigma_k\hat\mu_k\Biggr)^{1-y_{D-1}}~,
\end{eqnarray}
where $m=n$ or $m=n-\hat\mu$ and $y_{D-1}$ 
($0\le y_{D-1}\le 1$) is the 
interpolation parameter and can be 
regarded as the coordinate of $c(n)$ 
in the $\hat\mu_{D-1}$ direction. 
We use (\ref{eq:x4}) to define the second 
interpolation in the $\hat\mu_{D-2}$ direction as
\begin{eqnarray}
  \label{eq:x5}
  && w^m\Biggl(n+\sum_{k=1}^{D-3}\sigma_k\hat\mu_k
  +y_{D-2}\hat\mu_{D-2}+y_{D-1}\hat\mu_{D-1}\Biggr) \nonumber\\
  &&\equiv w^m\Biggl(n+\sum_{k=1}^{D-3}
  \sigma_k\hat\mu_k+\hat\mu_{D-2}
  +y_{D-1}\hat\mu_{D-1}\Biggr)^{y_{D-2}}
  w^m\Biggl(n+\sum_{k=1}^{D-2}\sigma_k\hat\mu_k
  +y_{D-1}\hat\mu_{D-1}\Biggr)^{1-y_{D-2}}~,\nonumber\\
\end{eqnarray}
where $y_{D-2}$ ($0\le y_{D-2}\le 1$) is the new parameter for the 
interpolation. 

We can carry out such interpolation 
procedure step by step until all the 
points in $p(n,\mu)$ are covered. 
After $l$ steps, we obtain
\begin{eqnarray}
  \label{eq:x6}
  && w^m\Biggl(n+\sum_{k=1}^{D-l-1}\sigma_k\hat\mu_k
  +\sum_{k=D-l}^{D-1}y_k\hat\mu_k\Biggr) \nonumber\\
  &&\hskip 2cm \equiv w^m\Biggl(n+\sum_{k=1}^{D-l-1}
  \sigma_k\hat\mu_k+\hat\mu_{D-l}
  +\sum_{k=D-l+1}^{D-1}y_k\hat\mu_k\Biggr)^{y_{D-l}} \nonumber\\
  &&\hskip 3cm \times w^m\Biggl(n+\sum_{k=1}^{D-l-1}\sigma_k\hat\mu_k
  +\sum_{k=D-l+1}^{D-1}y_k\hat\mu_k\Biggr)^{1-y_{D-l}}~.
\end{eqnarray}
In this way we can construct the parallel 
transport matrix $w^m(x)$ as 
\begin{eqnarray}
  \label{eq:x7-2}
  w^m(x)=\prod_{\{\sigma_k=0,1\}_{k=1,\cdots,D-1}}
  w^m\Biggl(n+\sum_{k=1}^{D-1}\sigma_k 
  \hat\mu_k\Biggr)^{\prod_{k=1}^{D-1}
    (\sigma_ky_k+(1-\sigma_k)(1-y_k))}
\end{eqnarray}
for any point $x\in p(n,\mu)$ given by
\begin{eqnarray}
  \label{eq:x7}
  x=n+\sum_{k=1}^{D-1}y_k\hat\mu_k~, \qquad (0\le y_k\le 1)~.
\end{eqnarray}
In deriving (\ref{eq:x7-2}) we have made a heavy use of the naive 
distribution law of exponents mentioned above. 
It is very important to note that the parallel transport function
$\{w^n(x)\}_{x\in \partial c(n)}$ defines a continuous mapping 
$\partial c(n)\rightarrow U(1)$ as can be verified directly from 
(\ref{eq:x7-2}).
 
In two and four dimensions $w^m(x)$ are explicitly 
given by
\begin{eqnarray}
  \label{eq:x7-3}
  D=2: ~~w^m(x)&=&w^m(n)^{1-y_1}w^m(n+\hat\mu_1)^{y_1}\\
  D=4: ~~w^m(x)&=&w^m(n)^{(1-y_1)(1-y_2)(1-y_3)}
  w^m(n+\mu_1)^{y_1(1-y_2)(1-y_3)}
  w^m(n+\hat\mu_2)^{(1-y_1)y_2(1-y_3)}\nonumber \\
  &&\times w^m(n+\hat\mu_3)^{(1-y_1)(1-y_2)y_3}
  w^m(n+\hat\mu_2+\hat\mu_3)^{(1-y_1)y_2y_3}\nonumber \\
  &&\times 
  w^m(n+\hat\mu_1+\hat\mu_3)^{y_1(1-y_2)y_3}
  w^m(n+\hat\mu_1+\hat\mu_2)^{y_1y_2(1-y_3)}\nonumber \\
  &&\times 
  w^m(n+\hat\mu_1+\hat\mu_2+\hat\mu_3)^{y_1y_2y_3}
\end{eqnarray}

Following ref. \cite{GKLSW}, we define the transition functions 
$v_{n,\mu}(x)$ interpolated over $p(n,\mu)$ by 
\begin{eqnarray}
  \label{eq:x8}
  v_{n,\mu}(x)\equiv w^{n-\hat\mu}(x)w^n(x)^{-1}~.
\end{eqnarray}
With this definition one can easily show that 
the transition functions 
indeed satisfy the cocycle conditions (\ref{eq:cocycle}).
It is not so difficult to compute the transition functions 
explicitly 
in arbitrary dimensions. We first note that the interpolated 
transition 
functions $v_{n,\mu}(x)$ ($x\in p(n,\mu)$) are given by
\begin{eqnarray}
  \label{eq:x9-2}
  v_{n,\mu}(x)=\prod_{\{\sigma_k=0,1\}_{k=1,\cdots,D-1}}
  v_{n,\mu}\Biggl(n+\sum_{k=1}^{D-1}\sigma_k 
  \hat\mu_k\Biggr)^{\prod_{k=1}^{D-1}
    (\sigma_ky_k+(1-\sigma_k)(1-y_k))}~.
\end{eqnarray}
Again use has been made of the naive manipulation on the exponents. 
For the corner points of $p(n,\mu)$ it is straightforward to 
show that the 
transition functions are given by
\begin{eqnarray}
  \label{eq:x9-3}
  v_{n,\mu}\Biggl(n+\sum_{k=1}^{D-1}\sigma_k\hat\mu_k\Biggr)
  =v_{n,\mu}(n)\exp\Biggl[
  i\sum_{{k=1 \atop \mu_k<\mu}}^{D-1}\sigma_k
  F_{\mu_k\mu}(n-\hat\mu+\sigma_1\hat\mu_1
  +\cdots+\sigma_{k-1}\hat\mu_{k-1})\Biggr]~,
\end{eqnarray}
where the field strengths are given by (\ref{eq:fieldstr}) and 
$v_{n,\mu}(n)=w^{n-\hat\mu}(n)w^n(n)^{-1}
=U_\mu(n-\hat\mu)$ is consistent with (\ref{eq:x8}). 
Putting this into the rhs of (\ref{eq:x9-2}), we obtain 
$v_{n,\mu}(x)$ 
explicitly as
\begin{eqnarray}
  \label{eq:x9-4}
  v_{n,\mu}(x)&=&v_{n,\mu}(n)\exp\Biggl[
  i\sum_{{k=1 \atop \mu_k<\mu}}^{D-1}
  \sum_{\{\sigma_j=0,1\}_{j=1,\cdots,k-1}}
  F_{\mu_k\mu}\Biggl(n-\hat\mu+\sum_{l=1}^{k-1}
  \sigma_l\hat\mu_l\Biggr)
  \nonumber \\
  &&\hskip 3cm \times\prod_{l=1}^{k-1}(\sigma_l y_l
  +(1-\sigma_l)(1-y_l))\:y_k
  \Biggr]
\end{eqnarray}
So far we have assumed that $|a_\mu(n)|$ is sufficiently 
small for any 
link variable so that all the mathematical manipulations 
concerning the 
distribution law of exponents can be justified. 
The final expression (\ref{eq:x9-4}), however, 
satisfies all the desired 
properties expected for the transition functions such as the gauge 
covariance and the cocycle conditions even for 
general configurations 
as far as the exceptional configurations \cite{phil} containing  
$F_{\mu\nu}(n)=\pm\pi$ for some $\mu$, $\nu$ and $n$ is excluded and 
the field strengths satisfy the Bianchi identities. Henceforth, we 
consider general field configurations not necessarily 
restricted to be 
small and regard (\ref{eq:x9-4}) as the definition of the transition 
functions. 

In the classical continuum limit we may retain only the 
leading terms as 
\begin{eqnarray}
  \label{eq:contlim}
  F_{\mu_k\mu}\Biggl(n-\hat\mu+\sum_{l=1}^{k-1}\sigma_l
  \hat\mu_l\Biggr)
  &=&F_{\mu_k\mu}(n)+\cdots
\end{eqnarray}
The transition function (\ref{eq:x9-4}) reduces to a simple 
form in the 
classical continuum limit as
\begin{eqnarray}
  \label{eq:x9-4-2}
  v_{n,\mu}(x)&=&v_{n,\mu}(n)
  \exp\Biggl[i\sum_{{k=1 \atop \mu_k<\mu}}^{D-1}
  F_{\mu_k\mu}(n)y_k+\cdots\Biggr]~.
\end{eqnarray}

Concrete expressions of the transition functions in some lower 
dimensions can be easily found form (\ref{eq:x9-4});
\begin{eqnarray}
  \label{eq:x9-5}
  D=2: && \nonumber\\
  v_{n,1}(x)&=& v_{n,1}(n)~, \qquad (x\in p(n,1))~, \nonumber \\
  v_{n,2}(x)&=& v_{n,2}(n){\rm e}^{iy_1F_{12}(n-\hat2)}~, \qquad 
  (x\in p(n,2))~, \\
  D=4: && \nonumber \\
  \label{eq:x9-6}
  v_{n,1}(x)&=& v_{n,1}(n)~, \qquad (x\in p(n,1))~, \nonumber \\
  v_{n,2}(x)&=& v_{n,2}(n){\rm e}^{iy_1F_{12}(n-\hat2)}~, \qquad 
  (x\in p(n,2))~, \nonumber \\
  v_{n,3}(x)&=& v_{n,3}(n){\rm e}^{i(y_1F_{13}(n-\hat3)
    +(1-y_1)y_2F_{23}(n-\hat3)+y_1y_2F_{23}(n+\hat1-\hat3))}~, 
  \qquad 
  (x\in p(n,3))~, \nonumber \\
  v_{n,4}(x)&=& v_{n,4}(n)\exp i[y_1F_{14}(n-\hat4)
  +(1-y_1)y_2F_{24}(n-\hat4)+(1-y_1)(1-y_2)y_3F_{34}(n-\hat4)
  \nonumber\\
  &&\hskip 2.2cm +y_1y_2F_{24}(n+\hat1-\hat4)
  +y_1(1-y_2)y_3F_{34}(n+\hat1-\hat4)
  \nonumber\\
  &&\hskip 2.2cm +(1-y_1)y_2y_3F_{24}(n+\hat2-\hat4)
  +y_1y_2y_3F_{34}(n+\hat1+\hat2-\hat4)]~, \nonumber \\
  &&\hskip 10cm (x\in p(n,4))~.
\end{eqnarray}
It is instructive to see in four dimensions that the Bianchi 
identities is necessary for the cocycle conditions 
(\ref{eq:cocycle}) 
to be fulfilled. Let us consider the case $\mu=2$ and $\nu=3$, then 
we have for $x\in p(n,2,3)$ 
\begin{eqnarray}
  \label{eq:concexcoc}
  && v_{n,2}(x)=v_{n,2}(n){\rm e}^{iy_1F_{12}(n-\hat2)}~, \qquad
  v_{n,3}(x)=v_{n,3}(n){\rm e}^{iy_1F_{13}(n-\hat3)}~, \nonumber\\
  && v_{n-\hat3,2}(x)=v_{n,2}(x)|_{y_3=1,n\rightarrow n-\hat3}
  =v_{n-\hat3,2}(n-\hat3){\rm e}^{iy_1F_{12}(n-\hat2-\hat3)}~, 
  \nonumber\\
  && v_{n-\hat2,3}(x)=v_{n,3}(x)|_{y_2=1,n\rightarrow n-\hat2}
  =v_{n-\hat2,3}(n-\hat2){\rm e}^{i(y_1F_{13}(n-\hat2-\hat3)
      +(1-y_1)F_{23}(n-\hat2-\hat3)
      +y_1F_{23}(n+\hat1-\hat2-\hat3))}~.
  \nonumber\\
\end{eqnarray}
The cocycle condition in the present case follows from the relations 
\begin{eqnarray}
  \label{eq:vvv}
  && v_{n-\hat2,3}(n-\hat2)v_{n,2}(n){\rm e}^{iF_{23}(n-\hat2-\hat3)}
  =v_{n-\hat3,2}(n-\hat3)v_{n,3}(n)~, \nonumber\\
  && {\it \Delta}_{[1}F_{23]}(n-\hat2-\hat3)=0~.
\end{eqnarray}

\section{Topological charge of lattice abelian gauge theory on 
the periodic lattice}
\label{sec:tclagtL}
\setcounter{equation}{0}

We have constructed a set of transition functions (\ref{eq:x9-4}) 
from the link variables on the lattice $\Lambda$. 
The transition functions in turn define a fiber bundle 
over $T^D$, for which the topological charge $Q$ can 
be computed unambiguously by (\ref{eq:9}). In order to 
compute $Q$, let us 
define 1-form $d\Lambda_{n,\mu}(x)$ on $p(n,\mu)$ by
\begin{eqnarray}
  \label{eq:x10}
  d\Lambda_{n,\mu}(x)=-i\:v_{n,\mu}(x)dv_{n,\mu}(x)^{-1}~,
\end{eqnarray}
where $d$ is the ordinary exterior derivative with respect to the 
continuous coordinates $x_\mu$. Then $Q$ is given by
\begin{eqnarray}
  \label{eq:x10-2}
  Q&=&\sum_{n\in\Lambda}q(n)~,
\end{eqnarray}
where $q$ is the topological charge density satisfying 
$q(n+L\hat\mu)=q(n)$ for $\mu=1,\cdots,D$ and can be chosen to be
\begin{eqnarray}
  \label{eq:x11}
    q(n)&=&(-1)^{\frac{N(N-1)}{2}}C_N
  \epsilon_{\mu_1\mu_2\cdots\mu_D}
  \int_{p(n,\mu_1,\mu_2,\cdots,\mu_N)}
  d^Nx
  \partial_{\mu_{N+1}}\Lambda_{n-\hat\mu_2-\cdots-\hat\mu_N,\mu_1} 
  \partial_{\mu_{N+2}}\Lambda_{n-\hat\mu_3-\cdots-\hat\mu_N,\mu_2} 
  \nonumber \\
  &&\hskip 9cm \times\cdots\times \partial_{\mu_D}\Lambda_{n,\mu_N}~.
\end{eqnarray}
From (\ref{eq:x9-4}) and (\ref{eq:x10}) one can see that $q(n)$ 
is invariant 
under the gauge transformations (\ref{eq:gtraa}), ultralocal 
in the gauge potential and a sum of products of $N$ field 
strengths. Furthermore, the classical continuum limit can be 
obtained from 
(\ref{eq:x9-4-2}) as
\begin{eqnarray}
  \label{eq:x11-2}
  q(n)=\frac{1}{2^N}C_N
  \epsilon_{\mu_1\nu_1\cdots\mu_N\nu_N}
  F_{\mu_1\nu_1}(n)\cdots F_{\mu_N\nu_N}(n)+\cdots~.
\end{eqnarray}
These properties together with the topological invariance of 
$Q$, i.e., 
the invariance under arbitrary local variations of the gauge 
potential 
as in (\ref{eq:topinv}), which is obvious by construction, in fact 
determine the structure of $q(n)$ by the theorem (\ref{eq:theorem}). 
In the present case we have $\beta_{\mu_1\nu_1\cdots\mu_l\nu_l}=0$ 
for 
$l<N$ and $\beta_{\mu_1\nu_1\cdots\mu_N\nu_N}=
C_N/2^N\:\epsilon_{\mu_1\nu_1\cdots\mu_N\nu_N}$ as can be seen from 
(\ref{eq:x11-2}). Furthermore, the current divergence term in 
(\ref{eq:theorem}) does not contribute to the topological charge $Q$ 
due to the periodic boundary conditions. 
We thus obtain the topological charge 
\begin{eqnarray}
  \label{eq:x11-4}
  Q&=&\frac{1}{2^N}C_N\sum_{n\in\Lambda}
  \epsilon_{\mu_1\nu_1\cdots\mu_N\nu_N}
  F_{\mu_1\nu_1}(n)F_{\mu_2\nu_2}(n+\hat\mu_1+\hat\nu_1) \nonumber\\
  &&\hskip 2cm\times\cdots\times
  F_{\mu_N\nu_N}(n+\hat\mu_1+\hat\nu_1+\cdots+\hat\mu_{N-1}
  +\hat\nu_{N-1})~.
\end{eqnarray}
The summand of this expression is nothing but the Chern character 
obtained 
solely from the noncommutative differential calculus on the 
lattice. This 
establishes the connection between the lattice analogue of the 
Chern 
character and the topology of the fiber bundle over the discrete 
lattice 
as in continuum theory.

It is possible to express the topological charge in terms of 
the magnetic 
fluxes $\phi_{\mu\nu}(x)$ through the $\mu\nu$-plane \cite{lus5}. 
They are
defined by 
\begin{eqnarray}
  \label{eq:17}
  \phi_{\mu\nu}(n)&=&\sum_{s,t=0}^{L-1}F_{\mu\nu}(n+s\hat\mu
  +t\hat\nu) 
  \nonumber\\
  &=&2\pi\sum_{s,t=0}^{L-1}n_{\mu\nu}(n+s\hat\mu+t\hat\nu) ~,
\end{eqnarray}
where use has been made of ({\ref{eq:ftoa}). We immediately see that 
they are gauge invariant and  integer multiples of $2\pi$. 
Furthermore, 
they can be shown to be constants by virtue of the 
Bianchi identities 
\cite{lus5}. Since the field strengths are continuous 
with respect to 
continuous changes of link variables as far as the exceptional gauge 
field configurations are excluded, the fluxes $\phi_{\mu\nu}$ 
must be  
invariant under such continuous changes of link variables. 
This implies that the sets of link variables 
$\{U_\mu(n)\}_{n\in\Lambda,\mu=1,\cdots,D}$ with distinct fluxes 
$\{\phi_{\mu\nu}\}_{\mu,\nu=1,\cdots,D}$ are topologically disjoint 
one another. 

For later convenience, we introduce a set of integers $m_{\mu\nu}
=-m_{\nu\mu}$ by 
\begin{eqnarray}
  \label{eq:18}
  m_{\mu\nu}=\sum_{s,t=0}^{L-1}n_{\mu\nu}(n+s\hat\mu+t\hat\nu)~.
\end{eqnarray}
If we define $\rho_{\mu\nu}(n)$ by
\begin{eqnarray}
  \label{eq:19}
  \rho_{\mu\nu}(n)&=&2\pi \Biggl(n_{\mu\nu}(n)
  -\frac{1}{L^2}m_{\mu\nu}\Biggr)~, 
\end{eqnarray}
then we have
\begin{eqnarray}
  \label{eq:20}
  \sum_{s,t=0}^{L-1}\rho_{\mu\nu}(n+s\hat\mu+t\hat\nu)=0~.
\end{eqnarray}
Furthermore, $\rho_{\mu\nu}(n)$ satisfies the Bianchi identities. 
Now using the lemma given in the next section, it is always possible 
to find a periodic vector field $\lambda_\mu$ on $\Lambda$ satisfying
\begin{eqnarray}
  \label{eq:21}
  {\it \Delta}_\mu\lambda_\nu(n)-{\it \Delta}_\nu \lambda_\mu(n)
  &=&\rho_{\mu\nu}(n)~.
\end{eqnarray}
This implies that the field strength $F_{\mu\nu}(n)$ can always 
be written as 
\begin{eqnarray}
  \label{eq:22}
  F_{\mu\nu}(n)&=&{\it \Delta}_\mu\tilde a_\nu(n)-{\it \Delta}_\nu 
  \tilde a_\mu(n)
  +\frac{2\pi}{L^2}m_{\mu\nu}~, 
\end{eqnarray}
where we have introduced a vector field on $\Lambda$ by 
$\tilde a_\mu(n)
\equiv a_\mu(n)+\lambda_\mu(n)$. It is possible to find a gauge 
potential $\tilde A_\mu(n)$ in terms of $\tilde a_\mu(n)$ and 
$m_{\mu\nu}$ as
\begin{eqnarray}
  \label{eq:gpot}
  \tilde A_\mu(n)&=&\tilde a_\mu(n)-\frac{2\pi}{L^2}
  \sum_{\nu>\mu}m_{\mu\nu}n_\nu~.
\end{eqnarray}
This satisfies the periodicity differing from (\ref{eq:aperio}) as 
\begin{eqnarray}
  \label{eq:aperio2}
  \tilde A_\mu(n+L\hat\nu)-\tilde A_\mu(n)&=&\cases{%
    -2\pi m_{\mu\nu}/L & $(\mu<\nu)$  \cr
    0                    & $(\mu\ge\nu)$}~.
\end{eqnarray}
Putting (\ref{eq:22}) into (\ref{eq:x11-4}) and noting that the 
periodic vector 
field $\tilde a_\mu(n)$ does not contribute to the topological 
charge, we obtain 
\begin{eqnarray}
  \label{eq:23}
  Q&=&\frac{1}{2^NN!}\epsilon_{\mu_1\nu_1\cdots\mu_N\nu_N}
  m_{\mu_1\nu_1}\cdots m_{\mu_N\nu_N}~,
\end{eqnarray}
where use has been made of the explicit form of $C_N$ given in 
(\ref{eq:5}). 
The rhs of this expression is manifestly an integer and 
topological invariant 
as it should be. 

\section{Lemma}
\label{sec:lemma}
\setcounter{equation}{0}

In this section we describe the Lemma used in the previous section. 
It turns out to be useful in finding link variables for given field 
strengths. We now state the lemma in the following form.
\vskip .3cm
\noindent
{\bf Lemma:} {\sl Let $\rho_{\mu\nu}$ be an antisymmetric 
tensor field 
on the $D$ dimensional periodic lattice $\Lambda$ satisfying
\begin{eqnarray}
  \label{eq:24}
  && {\it \Delta}_{[\lambda}\rho_{\mu\nu]}(n)=0~, \qquad
  \rho_{\mu\nu}(n+L\hat\lambda)=\rho_{\mu\nu}(n)~, \nonumber \\
  &&\sum_{s,t=0}^{L-1}\rho_{\mu\nu}(n+s\hat\mu+t\hat\nu)=0~,
  \qquad (\lambda,\mu,\nu=1,\cdots,D)
\end{eqnarray}
then there exists a vector fields $\lambda_\mu$ satisfying 
\begin{eqnarray}
  \label{eq:25}
  {\it \Delta}_\mu\lambda_\nu(n)-{\it \Delta}_\nu\lambda_\mu(n)
  =\rho_{\mu\nu}(n)~, \qquad
  \lambda_\mu(n+L\nu)=\lambda_\mu(n)~, 
  \qquad (\mu,\nu=1,\cdots,D)~.
\end{eqnarray}}

\vskip .3cm\noindent
{\sl Proof:} The proof will proceed as in the case of 
solving (\ref{eq:dmmu}) for $m_\mu(n)$. In the present case 
$\lambda_\mu(n)$ must be periodic in the lattice coordinates. 
This makes the solution somewhat complicated.

We first assume the existence of $\lambda_\mu$ satisfying 
(\ref{eq:25}) and note that the Bianchi identity 
and the periodicity in (\ref{eq:24}) imply the relation
\begin{eqnarray}
  \label{eq:26}
  {\it \Delta}_\mu\Biggl(\sum_{s=0}^{L-1}
  \rho_{\lambda\nu}(n+s\hat\lambda)\Biggr)
  ={\it \Delta}_\nu\Biggl(\sum_{s=0}^{L-1}
  \rho_{\lambda\mu}(n+s\hat\lambda)\Biggr)~.
\end{eqnarray}
Now let us denote a sum of vector field $V_\mu$ along a lattice 
path $C$ from $a\in\Lambda$ to $b\in\Lambda$ through the lattice 
points $a+\hat\mu$, 
$a+\hat\mu+\hat\nu$, $\cdots$, $b-\hat\rho$ in this order by 
\begin{eqnarray}
  \label{eq:lli}
  \sum_CV_\mu(y){\it \Delta}y_\mu&=&V_\mu(a)+V_\nu(a+\hat\mu)
  +\cdots+V_\rho(b-\hat\rho)~,
\end{eqnarray}
then (\ref{eq:26}) guarantees that the vector field 
$\alpha_\lambda$ defined by the sum along any lattice path $C$ from 
$y=0$ to $y=n$ 
\begin{eqnarray}
  \label{eq:27}
  \alpha_\lambda(n)=\alpha_\lambda(0)+\sum_{C}\sum_{s=0}^{L-1}
  \rho_{\lambda\mu}(y+s\hat\lambda){\it \Delta}y_\mu
\end{eqnarray}
is independent of the path chosen. 
It is also periodic as can be seen from the third property of 
(\ref{eq:24}). From (\ref{eq:25}) and (\ref{eq:27}) we get
\begin{eqnarray}
  \label{eq:28}
  {\it \Delta}_\mu\Biggl(\sum_{s=0}^{L-1}
  \lambda_\nu(n+s\hat\nu)\Biggr)
  =\sum_{s=0}^{L-1}\rho_{\mu\nu}(n+s\hat\nu)
  =-{\it \Delta}_\mu\alpha_\nu(n)~.
\end{eqnarray}
Hence we may choose $\alpha_\mu(n)$ without loss of generality 
to satisfy
\begin{eqnarray}
  \label{eq:28-2}
  \alpha_\mu(n)=-\sum_{s=0}^{L-1}\lambda_\mu(n+s\hat\mu)
\end{eqnarray}
We next define $\psi_\mu(n)$ and $\omega_{\mu\nu}(n)$by
\begin{eqnarray}
  \label{eq:29}
  \psi_\mu(n)=\lambda_\mu(n)+\frac{1}{L}\alpha_\mu(n)~, 
  \qquad
  \omega_{\mu\nu}(n)=\rho_{\mu\nu}(n)
  +\frac{1}{L}({\it \Delta}_\mu\alpha_\nu(n)
  -{\it \Delta}_\nu\alpha_\mu(n))~.
\end{eqnarray}
From (\ref{eq:25}) and (\ref{eq:28-2}) one can easily show that 
these fulfill the following relations 
\begin{eqnarray}
  \label{eq:30}
  && {\it \Delta}_\mu\psi_\nu(n)-{\it \Delta}_\nu\psi_\mu(n)
  =\omega_{\mu\nu}(n)~, \qquad 
  {\it \Delta}_{[\lambda}\omega_{\mu\nu]}(n)=0~, \nonumber\\
  && \sum_{s=0}^{L-1}\psi_\mu(n+s\hat\mu)
  =\sum_{s=0}^{L-1}\omega_{\mu\nu}(n+s\hat\mu)=0~.
\end{eqnarray}
It is now possible to solve these with respect to $\psi_\mu(n)$ 
just like (\ref{eq:dmmu}) by working with the axial gauge 
$\psi_D(n)=0$. In this gauge $\psi_\mu(n)$ ($\mu=1,\cdots,D-1$) 
are given by
\begin{eqnarray}
  \label{eq:31}
  \psi_\mu(n)=-\sum_{y_D=0}^{n_D-1}
  \omega_{\mu D}(n_1,\cdots,n_{D-1},y_D)
  +\psi_\mu(n_1,\cdots,n_{D-1},0)~,
\end{eqnarray}
where $\psi_\mu(n_1,\cdots,n_{D-1},0)$ is periodic in the lattice 
coordinates and still to be determined. It must satisfy
\begin{eqnarray}
  \label{eq:32}
  && {\it \Delta}_\mu\psi_\nu(n_1,\cdots,n_{D-1},0)
  -{\it \Delta}_\nu\psi_\mu(n_1,\cdots,n_{D-1},0)
  =\omega_{\mu\nu}(n_1,\cdots,n_{D-1},0)~, \nonumber \\
  &&\sum_{n_\mu=0}^{L-1}\psi_\mu(n_1,\cdots,n_{D-1},0)=0~.
\end{eqnarray}
These are the equations (\ref{eq:30}) restricted to $n_D=0$, and 
can be solved again by choosing 
$\psi_{D-1}(n_1,\cdots,n_{D-1},0)=0$. 
This procedure can be continued until all the 
$\psi_\mu(n)$ are found.
This completes the proof of the existence of $\psi_\mu$ and, hence, 
$\lambda_\mu$ by (\ref{eq:29}).  

\vskip .3cm
To illustrate all these ideas in a concrete example, 
let us consider a 
constant field\footnote{General constant fields of 
arbitrary magnetic 
fluxes are explicitly given in ref. \cite{lus5}.} 
$F_{\mu\nu}(n)=\epsilon_{\mu\nu}B$ 
in two dimensions.\footnote{In two dimensions the Bianchi identities
${\it \Delta}_{[\lambda}F_{\mu\nu]}=0$ are trivially 
satisfied for any 
anti-symmetric tensor fields.} If we parameterize the link 
variables as in (\ref{eq:1}), 
$a_\mu(n)$ ($\mu=1,2$) satisfy
\begin{eqnarray}
  \label{eq:amu}
  F_{12}(n)={\it \Delta}_1 a_2(n)-{\it \Delta}_2 a_1(n)
  +2\pi n_{12}(n)=B~, \quad
  (-\pi\le a_\mu(n)<\pi~,~~ |n_{12}(n)|\le2)~.
\end{eqnarray}
The magnetic flux $\phi\equiv2\pi\phi_{12}$ is given by 
\begin{eqnarray}
  \label{eq:flix}
  \phi=BL^2=\sum_{n\in\Lambda}2\pi n_{12}(n)~.
\end{eqnarray}
This implies that $BL^2$ must be an integer multiple of $2\pi$. 
For $BL^2=2\pi$ we may choose $n_{\mu\nu}(n)$ as
\begin{eqnarray}
  \label{eq:nmn}
  n_{\mu\nu}(n)&=&\delta_{\tilde n_1,[L/2]}
  \delta_{\tilde n_2,[L/2]}~,
\end{eqnarray}
where we have introduced periodic lattice coordinates 
\begin{eqnarray}
  \label{eq:ntilde}
  \tilde n_\mu=n_\mu-L\epsilon\Biggl(\frac{n_\mu}{L}
  +\frac{1}{2}\Biggr)~, \qquad
  ([-L/2]<\tilde n_\mu
  \le [L/2])
\end{eqnarray}
with $\epsilon(x)$ being the stair-step function defined by
\begin{eqnarray}
  \label{eq:11-2}
  \epsilon(x)=n \quad \hbox{for}\quad n< x\le n+1~, \quad 
  (n\in{\bf Z})~.
\end{eqnarray}
The $a_\mu(n)$'s now satisfy the following equation 
\begin{eqnarray}
  \label{eq:damn}
  {\it \Delta}_1 a_2(n)-{\it \Delta}_2 a_1(n)
  =B-2\pi n_{12}(n)
  =B-2\pi\delta_{\tilde n_1,[L/2]}\delta_{\tilde n_2,[L/2]}~.
\end{eqnarray}
We can solve this equation by following the procedure 
described in the proof 
of the lemma. We thus obtain
\begin{eqnarray}
  \label{eq:a1a2}
  a_1(n)=-\frac{2\pi}{L}\tilde n_2\delta_{\tilde n_1,[L/2]}~, \qquad
  a_2(n)=\frac{2\pi}{L^2}\tilde n_1~. 
\end{eqnarray}
The gauge potential $A_\mu(n)$ can be found by solving 
(\ref{eq:dmmu}) 
in the present case. The $m_\mu(n)$'s are easily obtained from 
(\ref{eq:mmusol}) as
\begin{eqnarray}
  \label{eq:mmn}
  m_1(n)=-\delta_{\tilde n_1,[L/2]}\epsilon\Biggl(\frac{n_2}{L}
  +\frac{1}{2}\Biggr)~, \qquad
  m_2(n)=0~.
\end{eqnarray}
In deriving this use has been made of the relation
\begin{eqnarray}
  \label{eq:de}
    {\it \Delta}_\mu\tilde n_\mu=1-L\delta_{\tilde n_\mu,[L/2]}~, 
    \quad (\hbox{no sum on}~\mu)~.
\end{eqnarray}
We thus find the gauge potential as
\begin{eqnarray}
  \label{eq:11-6}
  A_1(n)=a_1(n)+2\pi m_1(n)=-LB\delta_{\tilde n_1,[L/2]}n_2~, 
  \qquad A_2(n)=B\tilde n_1~.
\end{eqnarray}
The nontrivial periodicity property of the gauge potential is 
\begin{eqnarray}
  \label{eq:11-7}
  A_1(n+L\hat2)-A_1(n)=-L^2B\delta_{\tilde n_1,[L/2]}
  =-2\pi\delta_{\tilde n_1,[L/2]}~.
\end{eqnarray}
It is also possible to write the gauge potential in the form 
(\ref{eq:gpot}) as 
\begin{eqnarray}
  \label{eq:tildeain2d}
  \tilde A_1(n)=-\frac{2\pi}{L^2}n_2=-Bn_2~, \qquad \tilde A_2(n)=0~.
\end{eqnarray}
Finally the topological charge is given by
\begin{eqnarray}
  \label{eq:16}
  Q=\frac{1}{4\pi}\sum_{n\in\Lambda}\epsilon_{\mu\nu}F_{\mu\nu}(n)
  =\frac{L^2B}{2\pi}=1~.
\end{eqnarray}

\section{Summary and discussion}
\label{sec:summary}

We have argued that the topological charge obtained from the lattice 
generalization of the Chern character is indeed an integer 
related to 
the winding number of a U(1) bundle constructed from the link 
variables 
by a smooth interpolation. It picks up the topological structure 
of the underlying lattice originating from the periodicity. 
The configuration space of the link variables that is topologically 
trivial and connected to the trivial one becomes topologically 
disjoint by excising the exceptional field configurations.
No two belonging to different connected components can be 
continuously deformed into each other without crossing the 
exceptional configurations where the topological charge may 
jump due to the discontinuities of the field strengths at the 
exceptional plaquette variables. Link variables with different 
magnetic fluxes belong to different connected components of 
the field configurations. We have established an explicit 
relation between the topological charge and the magnetic 
fluxes. 

The construction presented in this paper is rather indirect. 
One might think that the interpolation to a smooth fiber 
bundle played an essential role and there might exist 
different interpolations leading to different topological 
charges. But such is not the case. The topological information 
of the underlying lattice is carried by the gauge potentials
and the value of the topological charge is completely unique 
for a given gauge field configuration. In order to see that 
the topological charge (\ref{eq:x11-4}) is indeed an integer 
given by (\ref{eq:23}) no interpolation to a smooth bundle 
is necessary. This way of understanding, however, makes the 
topological meaning of (\ref{eq:x11-4}) obscure. So it is more 
desirable to have a formalism of topological invariants of 
fiber bundles over discrete lattices without leaning upon the 
interpolation method. 

What is the implication of the present analysis on the index theorem 
of the Ginsparg-Wilson Dirac operator on the lattice? We infer that 
the topological charge (\ref{eq:x11-4}) coincides with the index of 
the Ginsparg-Wilson Dirac operator on the periodic lattice $\Lambda$ 
as the index theorem suggests. In the case of infinite lattice the 
chiral anomaly coincides with the Chern character. However, both the 
index and the topological charge are not well-defined in general. 
On the other hand both of them are well-defined in the case of 
finite 
lattice. Unfortunately, we have very little known about the precise 
connection between them. The situation gets even worse 
for nonabelian theories.\footnote{As a related study, see 
ref. \cite{adam}.} 
The topological charges obtained so far in 
the literature by interpolations \cite{lus4,PS,GKLSW} are too 
complicated to manipulate 
and no differential geometric construction of Chern characters 
leading to topological invariants seems to be available on the 
lattice. 
We must extend the theorem given in the Introduction to nonabelian 
theories in order to classify the topological invariants on the 
lattice. It is a challenging problem to establish the lattice 
extension of the index theorem. 

\vskip .5cm
One of us (K.W.) is very grateful to
F. Sakata for the warm hospitality extended to him and to Faculty of
Science of Ibaraki University for the financial support during his
visit at Ibaraki University.


\end{document}